\documentclass[amsmath,aps]{revtex4-2}
\usepackage{graphicx}
\usepackage{dcolumn}
\usepackage{bm}
\usepackage{multirow} 
\usepackage[colorlinks=true, linkcolor=blue, citecolor=red, urlcolor=blue]{hyperref}

\begin{document}


\title{On damping of Rabi oscillations in two-photon Raman excitation in cold $^{87}$Rb atoms}

\author{Vijay Kumar$^{a,b}$}
\email{vijaykr@rrcat.gov.in}
\thanks{Corresponding author}
\author{S. P. Ram$^a$}
\author{S. Singh$^a$}
\author{Kavish Bhardwaj$^a$}
\author{V. B. Tiwari$^{a,b}$}
\author{S. R. Mishra$^{a,b}$}
\affiliation{$^a$Laser Physics Applications Division, Raja Ramanna Centre for Advanced Technology, Indore-452013, India}
\affiliation{$^b$Homi Bhabha National Institute, Training School Complex, Anushakti Nagar, Mumbai-400094, India}

\date{\today}

\begin{abstract}
Two-photon Raman excitation between the ground hyperfine states $|5 \ ^2S_{1/2}, F = 2\rangle$ and  $|5 \ ^2S_{1/2}, F = 1\rangle$ of $^{87}$Rb atom has been experimentally studied. The Rabi coupling strengths of various transition involved have been calculated in presence of a weak magnetic field. A density matrix formalism has been developed to understand the experimentally observed damping of Rabi oscillations of population in a hyperfine state of $^{87}$Rb atom during interaction with the Raman laser beams. The observed damping of Rabi oscillations has been attributed to the dephasing during the light atom interaction.
\end{abstract}

                           
\maketitle

\section{Introduction}
\noindent
The interaction of an atomic ensemble with an external electromagnetic (EM) field may lead to temporal oscillations of the population between two quantum states of the atom. These oscillations are known as Rabi oscillations \cite{kosugi2005theory, foot2005atomic}. The Rabi oscillations are dependent on the strength of input EM field and the detuning of the field frequency from the resonance frequency of the two states. The manipulation of population in different states using Rabi oscillations has applications in quantum computing \cite{quantum_computing,qiao2015rabi}, precision measurement \cite{precision_G,high_precision}, quantum optics \cite{scully1997quantum,quantum_optics}, etc. In a multilevel atomic system \cite{multi_level,sargent1976three,Manipulation_STIRAP}, the dynamics of Rabi oscillations is more complex to understand. For example, in three-level atomic systems, the stimulated two-photon Raman transition \cite{bateman2010stimulated,bernard2022atom,johnson2008rabi,Ramsey_BEC} is a technique for population manipulation between two states via an intermediate state.

There are various decoherence mechanisms that play an important role in the coherent excitation of the atoms  \cite{analysis_Rydberg,laser_linewidth,decoherence_ion_trap,decoherence_magnetic_field,scheme,top_hat,comparison_of_Gaussian}, which have effect on the Rabi oscillations. These include spontaneous emission  \cite{analysis_Rydberg} and dephasing (phase relaxation) \cite{damped,spontaneous_STIRAP}. The dephasing can be homogeneous or inhomogeneous \cite{analysis,theory_of_coherent_atomic_excitation}. The homogeneous dephasing includes the dephasing due to spontaneous emission, laser phase noise \cite{analysis_Rydberg}, fluctuations in the laser frequency and intensity \cite{laser_linewidth,decoherence_ion_trap}, and fluctuations in the magnetic field \cite{decoherence_magnetic_field}. In contrast to homogeneous dephasing, the inhomogeneous dephasing occurs due to Gaussian distribution of the atomic cloud \cite{scheme}, the Gaussian profile of Raman laser beams \cite{top_hat}, the misalignment of the interacting laser beam \cite{comparison_of_Gaussian}, and Doppler broadening in the atomic ensemble \cite{analysis_Rydberg}. The time scale on which Rabi oscillations appear must be smaller than the decoherence time \cite{fox2006quantum}.

In this work, we have investigated the Rabi oscillations of population in hyperfine states during the two-photon Raman excitation of $^{87}$Rb atom from the ground hyperfine states $|5 \ ^2S_{1/2}, F = 2\rangle$ to $|5 \ ^2S_{1/2}, F = 1\rangle$. A density matrix formulation has been used to model this excitation process and explain the experimentally observed damping of Rabi oscillations. The Lindblad master equation \cite{spontaneous_STIRAP,STA} has been solved by incorporating the decoherence effects due to spontaneous emission and dephasing. It is shown that the dephasing of atomic levels due to various incoherent processes has the dominant contribution to the observed damping of Rabi oscillations.
\section{Experimental}
\begin{figure}[h!]
\includegraphics[width=0.35\textwidth]{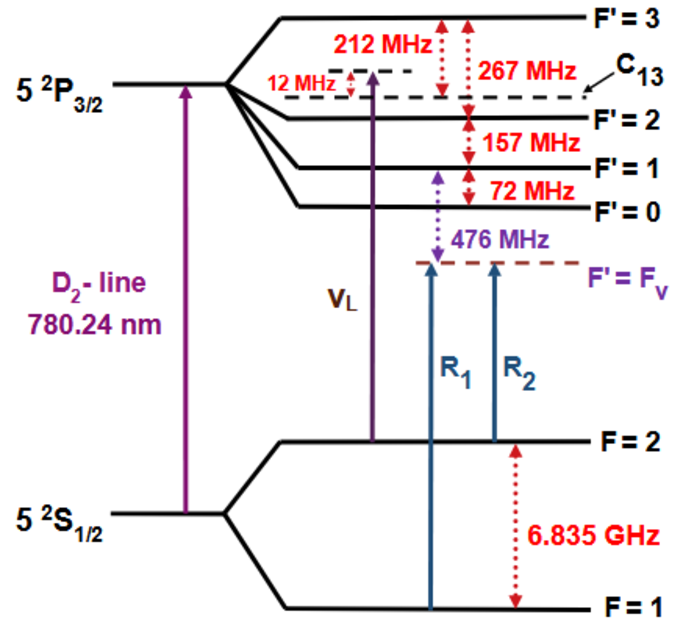}
\caption{\label{fig:Energy level SRT} Relevant energy levels of $^{87}$Rb atom for Raman transition.}
\end{figure}

\noindent
The laser cooled $^{87}$Rb atoms from a magneto-optical trap (MOT) were launched in vertical upward direction in an atomic fountain geometry using the moving molasses technique \cite{singh}. The experiments were performed on cold $^{87}$Rb atoms in $|5 \ ^2S_{1/2}, F = 2\rangle$ ground hyperfine state. The temperature of atom cloud before launch was $\sim$ 40 $\mu$K and number of atoms was  $\sim 5\times 10^7$. To generate a pair of Raman beams, a diode laser system was locked at frequency $\nu_L$, which was $\sim$ 12 MHz blue detuned from the peak of the cross-over transition $C_{13}$ in the D2-line of the $^{87}$ Rb atom. Using this laser beam, an electro-optic phase modulator (EOM) operating at $\nu_{RF} =6.835$ GHz was used to generate the frequencies $\nu_L$ and $\nu_L+ \nu_{RF}$. The output from the EOM was passed through an acousto-optic modulator (aligned in double-pass configuration) operating at a frequency of 350 MHz. This resulted in the generation of emissions at frequencies as $\nu_L$ + $\nu_{RF}$ - 700 MHz and $\nu_L$ - 700 MHz, which served as Raman beams ($R_1$ and $R_2$). In this arrangement, a frequency detuning of 476 MHz from $|5 \ ^2P_{3/2}, F' = 1\rangle$ level was achieved for both the Raman beams, as shown in Fig.~\ref{fig:Energy level SRT}. Both the Raman beams were linearly polarized and with polarization direction perpendicular to each other.

 The motivation for this experiment was to estimate the Rabi frequency for two-photon Raman transition, which is required for estimation of duration of $\pi$-pulse for Raman pulse atom interferometry. In the experiments, we have used linearly polarized counter-propagating Raman beams with their polarizations in mutually perpendicular directions. The Raman beams were applied on the atom cloud after $\sim$ 5 ms of the launching in vertical upward direction. An absorption probe beam, resonant to cooling transition of $^{87}$Rb atom, was aligned horizontally at a height of $\sim$ 9 cm from the MOT center in the path of atoms moving in the upward direction. The absorption of this probe beam was recorded to know the population in $F=2$ hyperfine state after applying the Raman beams pulse. The probe absorption signal (PAS) was recorded for different values of pulse duration of the Raman beams and for three different values of total power (7.5 mW, 9.5 mW and 11.5 mW) in the Raman beams. The results of these measurements are shown in section~\ref{sec:Results and discussion}. 
 
\section{Theoretical Model} 
\begin{figure}[h!]
\includegraphics[width=0.60\textwidth]{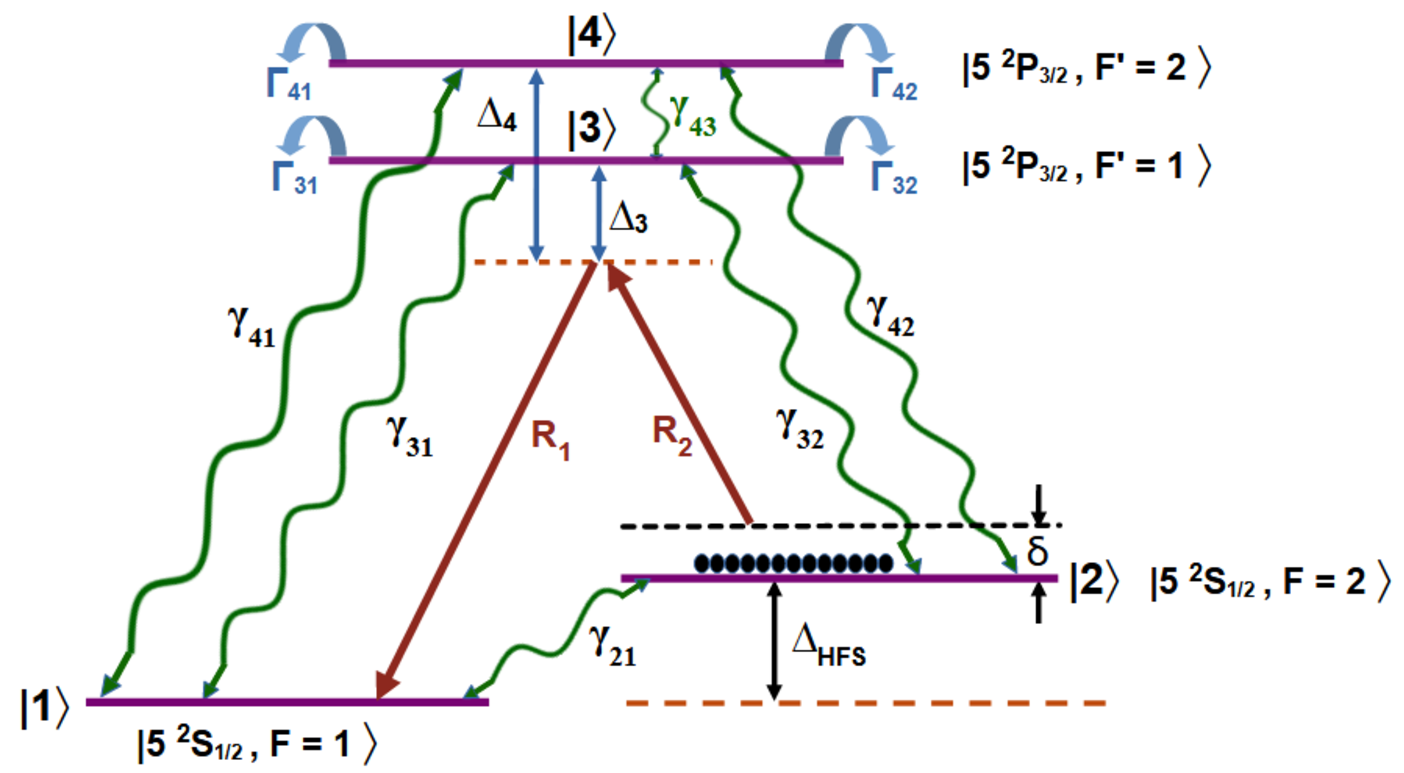}
\caption{\label{fig:SRT} Schematic diagram of partial energy levels of $^{87}Rb$ atom for two-photon Raman transition between two states $|2\rangle$ and $|1\rangle$. $\Delta_{HFS}$ is the hyperfine energy splitting  between states $|1\rangle$ and $|2\rangle$.}
\end{figure}

\noindent
The two-photon Raman transition between two ground hyperfine states is a frequently used phenomenon in the atom interferometry for precision measurement \cite{g_measurement,Zhou,theoreticalSRT}. We have modeled this Raman transition in $^{87}Rb$ atom by using a four-level scheme and solving the corresponding Lindblad master equation. Fig.~\ref{fig:SRT} shows the schematic of the energy levels of $^{87}Rb$ atom involved in two-photon stimulated Raman transition between states $|5 \ ^2S_{1/2}, F = 2\rangle$ and  $|5 \ ^2S_{1/2}, F = 1\rangle$. Two laser beams $R_2$ and $R_1$, having optical frequency $\omega_{2}$ and $\omega_{1}$ couple the two hyperfine levels $|2\rangle$ (i.e. $|5 \ ^2S_{1/2}, F = 2\rangle$) and $|1\rangle$ (i.e. $|5 \ ^2S_{1/2}, F = 1\rangle$) via intermediate states $|i\rangle$, with the corresponding coupling strengths $\Omega_{2i}$ and $\Omega_{i1}$ with $i = 3,4$. The single-photon detuning values $\Delta_3$ and $\Delta_4$ of the Raman beams from the excited states $|3\rangle$ and $|4\rangle$, are much larger than the linewidths of the levels $|3\rangle$ and $|4\rangle$. The two-photon detuning between Raman beams and states $|1\rangle$ and $|2\rangle$ is denoted by $\delta$. The parameters $\Gamma_{ij}$ represent spontaneous emission rates and $\gamma_{ij}$ (with $i \neq j$) represent the dephasing rates between the levels (Fig.~\ref{fig:SRT}). 

If atoms are initially in the state $|2\rangle$, then the stimulated Raman transition to the state $|1\rangle$ will involve absorption of a photon from one Raman beam ($R_2$) and the stimulated emission of a photon into the other Raman beam ($R_1$). For this two-photon stimulated Raman transition process, the Lindblad master equation \cite{Ultra-high_Lindblad,Spontaneous_Emission,Dephasing_Matrix} for evolution of the density matrix is given by,
\begin{equation}
 \frac{d\rho}{dt} = -i[H, \rho] + L_{\text{sp}}(\rho) + L_{\text{de}}(\rho) ,   
 \label{eq:Lindblad_Equation}
\end{equation}
where, $\rho$ is the density matrix and H is the total Hamiltonian. $L_{sp}(\rho)$ is the Lindblad term for spontaneous emission from the excited states and $L_{de}(\rho)$ is the Lindblad term for dephasing between different levels due to various processes. The Hamiltonian for the four-level $\Lambda$-system in Fig.~\ref{fig:SRT} under the interaction picture and rotating wave approximation (RWA) is given by \cite{bateman2010stimulated,Optical_Quantum_Information},
 \begin{equation}
    H = \begin{pmatrix}
   0 & 0 & \frac{\Omega_{31}}{2} & \frac{\Omega_{41}}{2} \\
   0 & -\delta & \frac{\Omega_{23}}{2} & \frac{\Omega_{24}}{2} \\
   \frac{\Omega_{31}}{2} & \frac{\Omega_{23}}{2} & \Delta_3 & 0 \\
   \frac{\Omega_{41}}{2} & \frac{\Omega_{24}}{2} & 0 & \Delta_4 \\
   \end{pmatrix} .  
   \label{eq:Hamiltonian}
 \end{equation}
Though the single-photon detuning $\Delta_3$ and $\Delta_4$ is large, there is still some probability of a single-photon transition, which can result in the transfer of population from the state $|2\rangle$ to $|3\rangle$ and $|4\rangle$. If the spontaneous decay rates $\Gamma_{31}$, $\Gamma_{32}$, $\Gamma_{41}$ and $\Gamma_{42}$ from the levels $|3\rangle$ and $|4\rangle$, then the Lindblad term for spontaneous emission is given by \cite{spontaneous_STIRAP,STA}, 
\begin{equation}
L_{\text{sp}}(\rho) = 
\left( \begin{array}{cc}
\Gamma_{31}\rho_{33} + \Gamma_{41}\rho_{44} & 0 \\
0 & \Gamma_{32}\rho_{33} + \Gamma_{42}\rho_{44} \\
-\frac{1}{2}(\Gamma_{31} + \Gamma_{32})\rho_{31} & -\frac{1}{2}(\Gamma_{31} + \Gamma_{32})\rho_{32} \\
-\frac{1}{2}(\Gamma_{41} + \Gamma_{42})\rho_{41}  & -\frac{1}{2}(\Gamma_{41} + \Gamma_{42})\rho_{42} \\
\end{array} \right.
\begin{aligned}
\left. \begin{array}{cc}
-\frac{1}{2}(\Gamma_{31} + \Gamma_{32})\rho_{13} & -\frac{1}{2}(\Gamma_{41} + \Gamma_{42})\rho_{14} \\
-\frac{1}{2}(\Gamma_{31} + \Gamma_{32})\rho_{23} & -\frac{1}{2}(\Gamma_{41} + \Gamma_{42})\rho_{24} \\
-(\Gamma_{31} + \Gamma_{32})\rho_{33} & -\frac{1}{2}(\Gamma_{31} + \Gamma_{32} + \Gamma_{41} + \Gamma_{42})\rho_{34} \\
-\frac{1}{2}(\Gamma_{31} + \Gamma_{32} + \Gamma_{41} + \Gamma_{42})\rho_{43} & -(\Gamma_{41} + \Gamma_{42})\rho_{44} \\
\end{array} \right)
\end{aligned}  
\label{eq:Spontaneous_Decay}
\end{equation} 
Similarly, the Lindblad term for dephasing \cite{Dephasing_Matrix, dephasing_tripod} between different levels can be written as, 
\begin{equation}
   L_{\text{de}}(\rho) = \begin{pmatrix}
   0 & -\gamma_{12}\rho_{12} & -\gamma_{13}\rho_{13} & -\gamma_{14}\rho_{14} \\
   -\gamma_{21}\rho_{21} & 0 & -\gamma_{23}\rho_{23} & -\gamma_{24}\rho_{24} \\
   -\gamma_{31}\rho_{31} & -\gamma_{32}\rho_{32} & 0 & -\gamma_{34}\rho_{34} \\
   -\gamma_{41}\rho_{41} & -\gamma_{42}\rho_{42} & -\gamma_{43}\rho_{43} & 0 
   \end{pmatrix} .
   \label{eq:Dephasing_Rates}
\end{equation}
\noindent
Using the above set of equations (Eqs.~(\ref{eq:Lindblad_Equation})–(\ref{eq:Dephasing_Rates})), we have numerically solved the Lindblad equation to evaluate the time-dependent populations and coherences for different levels.

In the presence of a finite stray magnetic field in an arbitrary direction, there are several possible pathways for two-photon Raman transition from state $|F=2\rangle$ to $|F=1\rangle$ involving Zeeman sublevels of all the four hyperfine states. Each pathway for Raman transition consists of two single-photon transitions, where an atom absorbs a photon from the one Raman beam and emits a photon into the other Raman beam by stimulated emission process. The details of such Raman transition routes are given in the \hyperref[sec:Appendix]{Appendix} at the end of this paper. 

Using the formalism (Eq.~(\ref{eq:Rabi_Frequency}) and Eq.~(\ref{eq:RMS_Rabi_Frequency}) discussed in \hyperref[sec:Appendix]{Appendix}, we can calculate the root mean square (rms) values of Rabi frequencies $\Omega_{23}$, $\Omega_{31}$, $\Omega_{24}$, and $\Omega_{41}$ for the single-photon transitions $|2\rangle\rightarrow|3\rangle$, $|3\rangle\rightarrow|1\rangle$, $|2\rangle\rightarrow|4\rangle$, and $|4\rangle\rightarrow|1\rangle$, respectively. We get the numerical values of $\Omega_{23}$, $\Omega_{31}$, $\Omega_{24}$, and $\Omega_{41}$ for a total Raman beams power (i.e. sum of power in both the Raman beams) of 9.5 mW as $2\pi \times 3.15$ MHz, $2\pi \times 5.25$ MHz, $2\pi \times 6.58$ MHz, and $2\pi \times 4.46$ MHz, respectively. Then  two-photon Rabi frequency for stimulated Raman transition $|2\rangle\rightarrow|1\rangle$ via levels $|3\rangle$ and $|4\rangle$ is given as \cite{AJDunningCoherent,kasevich1992}, 
\begin{equation}
    \Omega_{R} = \sum_i \frac{\Omega_{2i} \Omega_{i1}}{2 \Delta_i} ,
    \label{eq:Two_Photon_Rabi_Frequency}
\end{equation}
where sum is over index $i$ ($i=3,4$), and $\Delta_i$ is the single-photon detuning of Raman beam from the intermediate level $|i\rangle$. The two-photon Rabi frequency $\Omega_{R}$ for Raman transition $|2\rangle\rightarrow|1\rangle$, as calculated for our theoretical model (using Eq.~(\ref{eq:Two_Photon_Rabi_Frequency})), is 2$\pi$ $\times$ 40.55 kHz for total Raman beam power of 9.5 mW. By knowing $\Omega_{R}$ and detuning $\delta$, the effective Rabi frequency becomes
\begin{equation}
    \Omega_{eff} = \sqrt{\Omega_R^2 + \delta^2} .
    \label{eq:Effective_Rabi_Frequency}
\end{equation}

\noindent
The spontaneous emission rate $\Gamma_{F',m'_{F}}^{F, m_{F}}$ from the state $|F',m'_{F}\rangle$ to $|F,m_{F}\rangle$ is given as \cite{Rb_blue_transition},
\begin{align}
\Gamma_{F',m'_{F}}^{F, m_{F}} \propto \ & (2J' + 1)(2F + 1)(2F' + 1) \cdot
\left\{
\begin{array}{ccc}
J' & 1 & J \\
F & I & F'
\end{array}
\right\}^2 
\cdot
\left(
\begin{array}{ccc}
F' & 1 & F \\
m'_{F} & m_{F} - m'_{F} & -m_{F}
\end{array}
\right)^2 .
\label{transition_probability}
\end{align}
\noindent
The net spontaneous emission rate ($\Gamma_{F'F}$ ) from the state $|F'\rangle$ to $|F\rangle$ is given as,
\begin{equation}
\Gamma_{F'F} =   \sum_{m_F', m_F} \Gamma_{F',m'_{F}}^{F, m_{F}}   \     ,
\label{gamma-se}
\end{equation}
 where sum is over all allowed transitions from all magnetic sublevels $m'_F$ of the hyperfine state $F'$ to the magnetic sublevels $m_F$ of the hyperfine state F. The total spontaneous emission rate from $F'$ is given as $\Gamma = \sum_{F} \Gamma_{F'F}$ \cite{Rb_blue_transition}. The calculated spontaneous emission rates $\Gamma_{31}$, $\Gamma_{32}$, $\Gamma_{41}$ and $\Gamma_{42}$, in our case, are $2 \pi \times 5.03$ MHz, $2 \pi \times 1.03$ MHz, $2 \pi \times 3.03$ MHz and $2 \pi \times 3.03$ MHz, respectively.

\section{Results and discussion}
\label{sec:Results and discussion}
\begin{figure}[b!]
\includegraphics[width=0.85\textwidth]{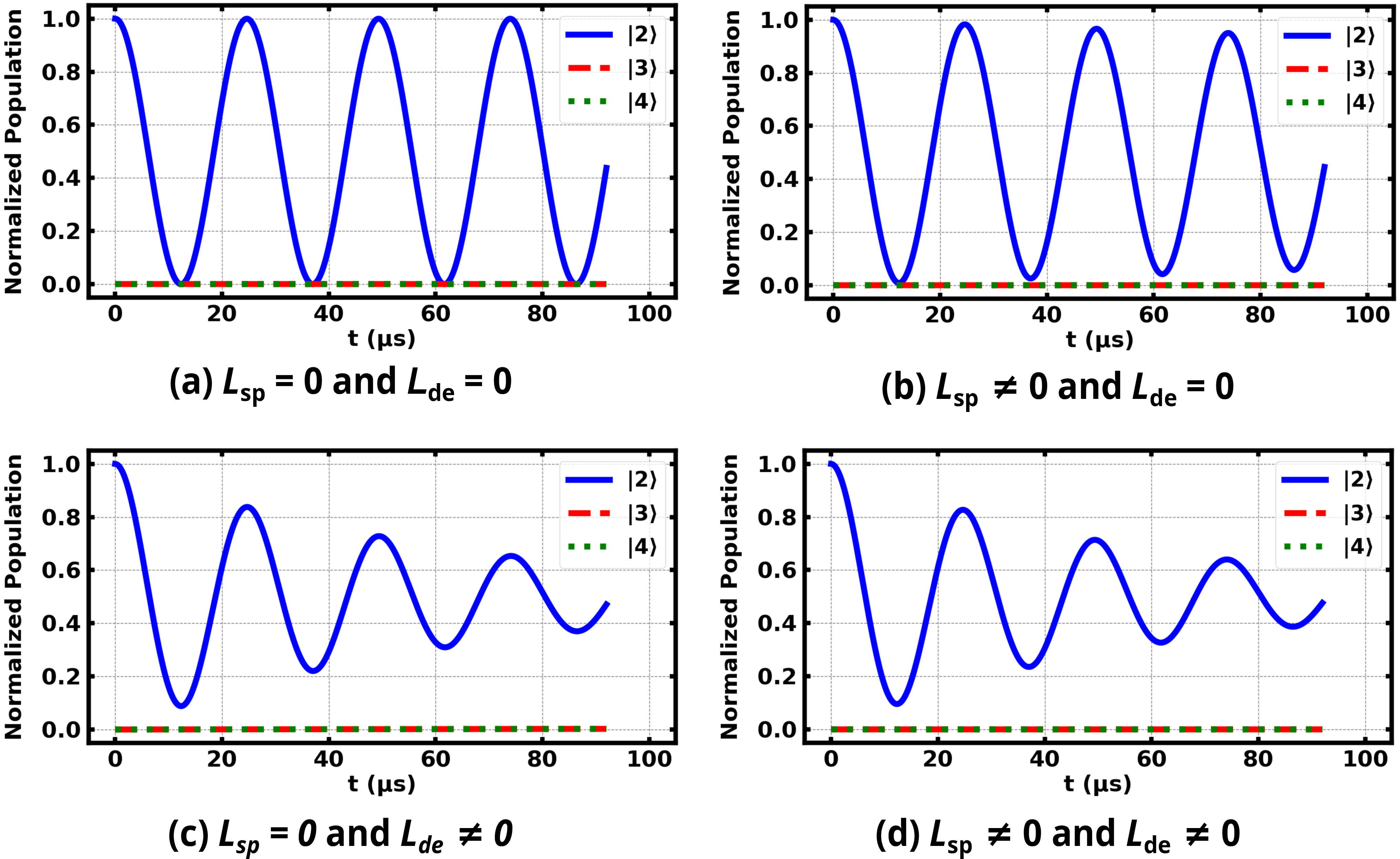}
\caption{\label{fig:Simulation Plot} The calculated value of population in level  $|5 ^2S_{1/2}, F=2\rangle$ of $^{87}Rb$ as function of time (t), for $\Omega_{23} = 2\pi \times 3.15$ MHz, $\Omega_{31} = 2\pi \times 5.25$ MHz, $\Omega_{24} = 2\pi \times 6.58$ MHz, $\Omega_{41} = 2\pi \times 4.46$ MHz and , $\Delta_3 = 2\pi \times 476$ MHz, $\Delta_4 = 2\pi \times 633$ MHz and $\delta = 0$. The dephasing and spontaneous emission parameter are described in the text.}
\end{figure}
\noindent
The theoretically calculated population in the state $|F=2\rangle$ after interaction of an atom with Raman beams for a time duration of `$t$' is shown in Fig.~\ref{fig:Simulation Plot}. The plot in Fig.~\ref{fig:Simulation Plot}(a) shows the Rabi oscillations of population without any decoherence mechanisms ($L_{sp} = L_{de} = 0$). The plot in Fig.~\ref{fig:Simulation Plot}(b) shows the effect of spontaneous emission ($L_{sp} \neq 0$ and $L_{de} = 0$) from the excited states with the values of $\Gamma_{ij}$ as calculated using Eq.~(\ref{gamma-se}). It is noted that effect of the spontaneous emission is very weak on the Rabi oscillations, as population in the excited states ($|3\rangle$ and $|4\rangle$) remains quite small ($\sim$ $10^{-5}$ to $10^{-3}$). This can be attributed to the large detuning of Raman beams from the excited states. 

Figure~\ref{fig:Simulation Plot}(c) shows the effect of dephasing ($L_{sp} = 0$ and $L_{de} \neq 0$) on the Rabi oscillations. The dephasing parameters used in the calculation are $\gamma_{ij}=\gamma_{ji} = 2\pi \times 0.1$ MHz  (for $i \in \{1,2\}, j \in \{3,4\}$), $\gamma_{12} = \gamma_{21} = 2\pi \times 5$ kHz and $\gamma_{34} = \gamma_{43} = 2\pi \times 6$ kHz. This plot shows that damping of Rabi oscillations is present due to effect of dephasing. Figure~\ref{fig:Simulation Plot}(d) shows the effect of spontaneous emission and dephasing both, with dephasing and spontaneous emission parameters same as in plots (c) and (b). In general, the damping of Rabi oscillations can occur due to spontaneous emission and dephasing, both, but we did not observe much contribution from the spontaneous emission (plots (b) and (d)). This is due to large detuning of Raman beams from the excited states.

\begin{figure*}[h!]
\centering
\includegraphics[width=1.0\textwidth]{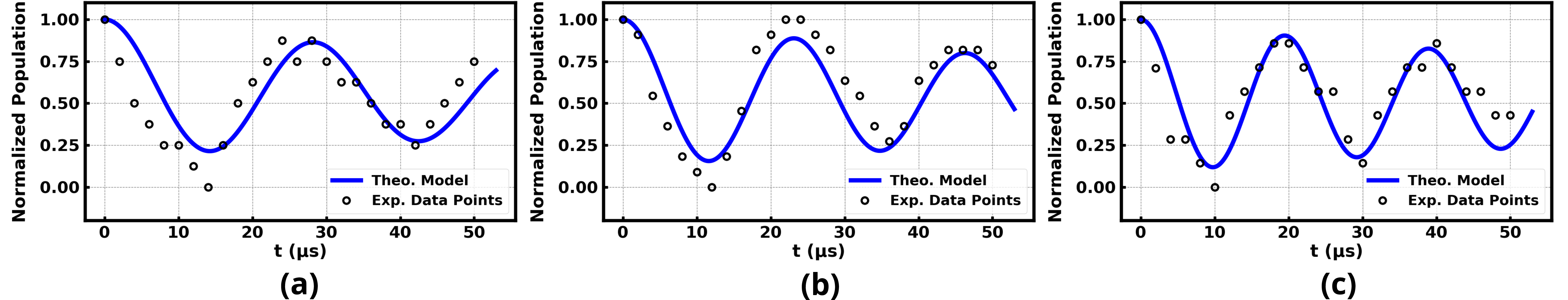}
    \caption{Rabi oscillations in atomic population of the ground hyperfine state $|5 \ ^2S_{1/2}, F = 2\rangle$ of the $^{87}Rb$ atom for total Raman beam power of (a) 7.5 mW (b) 9.5 mW, and (c) 11.5 mW, respectively. Black open circles represent the experimental data points and solid blue curves show the calculated values fit to the experimental data. The other parameters used in this simulation are the dephasing rates: $\gamma_{ij} = \gamma_{ji}= 2\pi \times 0.2$ MHz $(i \in \{1,2\} \ $and$ \ j \in \{3,4\}$), $\gamma_{12} = \gamma_{21} = 2\pi \times 2.4$ kHz,  and $\gamma_{34} = \gamma_{43} = 2\pi \times 5.0$ kHz. The two-photon Raman detuning ($\delta$) is taken as $2\pi \times 15$ kHz for each simulated curve.}
    \label{fig:Rabi_Oscillation}
\end{figure*}

The experimentally observed results on Rabi oscillations of population in F=2 hyperfine state under two-photon Raman excitation are shown in Fig.~\ref{fig:Rabi_Oscillation} for total Raman beams power of 7.5 mW, 9.5 mW, and 11.5 mW, respectively. The continuous curves in this figure show the calculated data for a closure agreement to the experimental results. For the calculated curves in this figure, the values of single-photon Rabi frequencies (rms values) $\Omega_{23}$, $\Omega_{31}$, $\Omega_{24}$, and $\Omega_{41}$ for three different total power of Raman beam (7.5 mW, 9.5 mW and 11.5 mW) are calculated using the data of \hyperref[sec:Appendix]{Appendix}. The RMS values of these single-photon Rabi frequencies for different Raman beams power are listed in Table \ref{Single-photon Rabi frequencies}. The spontaneous decay rates $\Gamma_{31}$, $\Gamma_{32}$, $\Gamma_{42}$, and $\Gamma_{41}$ are same as used in Fig.~\ref{fig:Simulation Plot}.
 
\begin{table*}[h!] 
\caption{Calculated rms values of single-photon Rabi frequencies between hyperfine levels of $^{87}$Rb atom for different total Raman beams power  ($P_{total}$).}
\scalebox{0.95}{ 
\renewcommand{\arraystretch}{1.2} 
\begin{tabular}{|c|c|c|c|c|}
  \hline
   \multicolumn{5}{|c|}{\textbf{Rabi frequencies for different total Raman beams power ($P_{total}$)}} \\
   \hline
   {\textbf{\boldmath{$P_{total}$}}} & \textbf{\boldmath{$\Omega_{23}/2\pi$ (MHz)}} & \textbf{\boldmath{$\Omega_{31}/2\pi$ (MHz)}} & \textbf{\boldmath{$\Omega_{24}/2\pi$ (MHz)}} & \textbf{\boldmath{$\Omega_{41}/2\pi$ (MHz)}}\\
   \hline
   7.5 mW & 2.80 & 4.67 & 5.86 & 3.97\\
   \hline
   9.5 mW & 3.15 &  5.25 & 6.58 & 4.46 \\
   \hline
   11.5 mW & 3.47  &  5.78 & 4.91 & 7.24 \\
   \hline
\end{tabular}
}
\label{Single-photon Rabi frequencies}
\end{table*}

From the Fig.~\ref{fig:Rabi_Oscillation}, it can be noted that experimentally observed Rabi frequencies for two photon Ram transitions are 2$\pi$ $\times$ 37.75 kHz, 2$\pi$ $\times$ 43.01 kHz, and 2$\pi$ $\times$ 51.94 kHz for Raman beams power of 7.5 mW, 9.5 mW, and 11.5 mW respectively. The corresponding theoretical values of effective Rabi frequency ($\Omega_{eff}$) for two-photon transition are 2$\pi$ $\times$ 35.45 kHz, 2$\pi$ $\times$ 43.24 kHz, and 2$\pi$ $\times$ 51.39 kHz for total Raman beam power of 7.5 mW, 9.5 mW, and 11.5 mW, respectively. These values show a good agreement between theory and experimental data. From the experimental data of Rabi ferquency, the estimated value of $\pi$-pulse durations are $13.25 \ \mu$s, $11.63 \ \mu$s, and $9.63 \ \mu$s for Raman beams power of 7.5 mW, 9.5 mW, and 11.5 mW, respectively. The damping behavior in the theoretical curves is due to finite value of dephasing parameters incorporated in the theoretical model in order to resemble the experimental observations. The dephasing in the experiments can arise due to several factors including intensity and linewidth fluctuations in the Raman beams, finite Zeeman splitting of levels, and finite Doppler broadening. We have not quantified the contribution of each of them to the dephasing values used in the theoretical calculations.

\section{Conclusion}
\noindent
 The Rabi oscillations in the population of the ground hyperfine state $|5 \ ^2S_{1/2}, F = 2\rangle$ of the $^{87}Rb$ atom have been studied under a two-photon Raman excitation process. Damping in the oscillations has been experimentally observed with an increase in the interaction duration of the Raman beams with atoms. By solving the Lindblad master equation, it has been shown that damping of Rabi oscillations in two-photon Raman transition from $|5 \ ^2S_{1/2}, F = 2\rangle$ to $|5 \ ^2S_{1/2}, F = 1\rangle$ is dominantly due to dephasing mechanism. The decoherence due to spontaneous emission has negligible contribution to the observed damping of Rabi oscillations. The sources of dephasing in the experiments could be the fluctuations in the intensity and linewidth of the Raman beams, finite Zeeman splitting of levels, and finite Doppler broadening.

\section*{Acknowledgments}
\noindent
We thank Shri Vivek Singh for insightful theoretical discussions. We are also thankful to Shri Sanjeev Bhardwaj for technical help during the experiments and to Shri Ayukt Pathak for the development of the controller system.



\newpage
\appendix
\renewcommand{\thefigure}{A.\arabic{figure}}
\setcounter{figure}{0} 
\renewcommand{\thetable}{A.\arabic{table}}
\setcounter{table}{0} 
\renewcommand{\theequation}{A.\arabic{equation}}
\setcounter{equation}{0} 
\section*{Appendix}
\label{sec:Appendix}
\begin{figure*}[h!]
    \centering
    \includegraphics[width=0.80\textwidth]{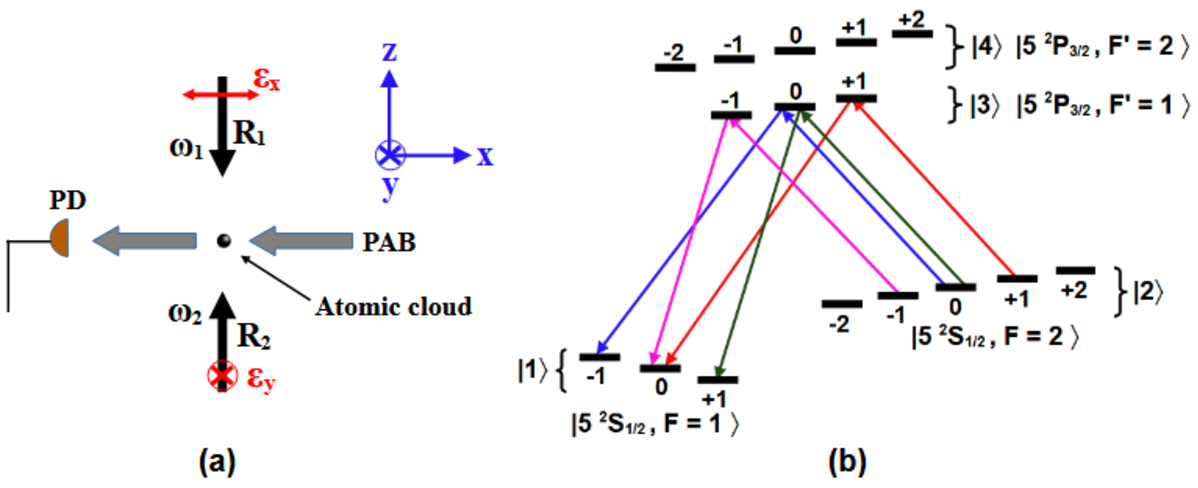}
    \caption{(a) shows the orientation of Raman beams. (b) shows different routes for two-photon Raman transition between ground hyperfine states (from $|F = 2\rangle$ to $ |F = 1\rangle$ ) via the intermediate state $|5 ^2P_{3/2}, F' = 1\rangle$, when $B_x$ component of magnetic field is considered.}
    \label{fig:figureA1}
\end{figure*}

\noindent
Figure ~\ref{fig:figureA1} shows two Raman beams $R_{1}$ and $R_{2}$ propagating in z-direction and having mutually orthogonal linear polarizations $\varepsilon_x$ and $\varepsilon_y$. The Raman beams interact with an atom in the presence of the stray magnetic field. Fig.~\ref{fig:figureA1}(a) shows the orientation of Raman beams and Fig.~\ref{fig:figureA1}(b) shows the various pathways for Raman transition from $|5 ^2S_{1/2}, F = 2\rangle$ to $|5 ^2S_{1/2}, F = 1\rangle$ via the intermediate state $|5 ^2P_{3/2}, F' = 1\rangle$ in presence of $B_x$ component of the magnetic field.

Since the stray magnetic field is oriented arbitrarily, each single-photon transition is governed by the direction of the polarization of the Raman beams and the direction of the stray magnetic field components ($B_x$, $B_y$, and $B_z$). The summary of possible single-photon transitions is given in Table \ref{Configuration of Raman Polarization}. Based on the direction of the magnetic field and polarization of Raman beams, the routes for Raman transitions from the state $|F = 2, m_F \rangle$ to $|F = 1, m_F \rangle$ via intermediate states $|F^{'}, m'_{F}\rangle$, with ($F^{'}=1,2$), are summarized in Table \ref{Raman Transition Routes due to stray magnetic field}.

\begin{table*}[h!]
   \caption{Single-photon transitions in an atom interacting with a linearly polarized laser beam (propagating along the $z$-direction) in the presence of a stray magnetic field. \label{tab:RamanPolarization}}
   \centering
   \renewcommand{\arraystretch}{1.7}  
   
   \begin{tabular}{|>{\centering\arraybackslash}p{2.8cm}|>{\centering\arraybackslash}p{2.8cm}|>{\centering\arraybackslash}p{4.5cm}|>{\centering\arraybackslash}p{1.5cm}|}
       \hline
        \textbf{Magnetic Field Component} & \textbf{Polarization of Laser Beam} & \textbf{Transition Type} & $\mathbf{\Delta m_F}$ \\ 
        \hline
        \multirow{2}{*}{$B_x$} & $\varepsilon_x$ & $\pi^0$ & $0$ \\ \cline{2-4}
                & $\varepsilon_y$ & $\pi^{+} = \frac{(\sigma^+ + \sigma^-)}{\sqrt{2}}$ & $\pm1$  \\ 
        \hline
        \multirow{2}{*}{$B_y$} &  $\varepsilon_x$  & $\pi^{+} = \frac{(\sigma^+ + \sigma^-)}{\sqrt{2}}$ & $\pm1$  \\ \cline{2-4}
                & $\varepsilon_y$ & $\pi^0$  & $0$  \\ 
        \hline   
        \multirow{2}{*}{$B_z$} &  $\varepsilon_x$  & $\pi^{+} = \frac{(\sigma^+ + \sigma^-)}{\sqrt{2}}$ & $\pm1$ \\ \cline{2-4}      
                &  $\varepsilon_y$  & $\pi^{-} = \frac{(\sigma^+ - \sigma^-)}{\sqrt{2}}$ & $\pm1$  \\ 
        \hline     
    \end{tabular}
    \label{Configuration of Raman Polarization}
\end{table*}

\begin{table*}[h!]
   \caption{Possible routes for the transition of a $^{87}$Rb atom from the ground hyperfine state $|F = 2\rangle$ to the ground hyperfine state $|F = 1\rangle$, when two Raman beams with mutually orthogonal linear polarizations ($R_1$ in the x-direction and $R_2$ in the y-direction) are used in a two-photon Raman transition in the presence of a magnetic field. Here, $F' \ (= 1,2) $ represents the hyperfine number of the intermediate level.}
   \centering
   \renewcommand{\arraystretch}{1.3} 

   \begin{tabular}{|>{\centering\arraybackslash}p{2.5cm}|>{\centering\arraybackslash}p{3.5cm}|>{\centering\arraybackslash}p{4.5cm}|>{\centering\arraybackslash}p{3.5cm}|}
         \hline
         \textbf{Magnetic Field Component} & \textbf{Transition Type} & \textbf{Two-Photon Raman Transition Routes} & \textbf{Via Intermediate State} \\ 
         \hline
         
         \multirow{2}{*}{$B_x$} & \multirow{2}{*}{$\pi^0 - \pi^+$} 
         & $|2,m_F\rangle \rightarrow |1,m_F + 1\rangle$ & $|F',m'_F = m_F\rangle$ \\  
         \cline{3-4}
         &  & $|2,m_F\rangle \rightarrow |1,m_F - 1\rangle$ & $|F',m'_F = m_F\rangle$ \\ 
         \hline
         
         \multirow{2}{*}{$B_y$} & \multirow{2}{*}{$\pi^+ - \pi^0$} 
         & $|2,m_F\rangle \rightarrow |1,m_F + 1\rangle$ & $|F',m'_F = m_F + 1\rangle$ \\  
         \cline{3-4}
         &  & $|2,m_F\rangle \rightarrow |1,m_F - 1\rangle$ & $|F',m'_F = m_F - 1\rangle$ \\ 
         \hline
         
         \multirow{2}{*}{$B_z$} & \multirow{2}{*}{$\pi^+ - \pi^-$} 
         & $|2,m_F\rangle \rightarrow |1,m_F\rangle$ & $|F',m'_F = m_F + 1\rangle$ \\  
         \cline{3-4}
         &  & $|2,m_F\rangle \rightarrow |1,m_F\rangle$ & $|F',m'_F = m_F - 1\rangle$ \\ 
         \hline
   \end{tabular}

   \label{Raman Transition Routes due to stray magnetic field}
\end{table*}
 
 The Rabi frequency for a single-photon transition from the state $|J, F, m_F \rangle$ to $|J', F', m'_F \rangle$, is calculated using the expression \cite{bateman2010stimulated,comparison_of_Gaussian,Daniel_A_steck_2003},
\begin{align}
\Omega_{F,m_{F}}^{F', m'_{F}} = & \ \frac{E}{\hbar} \times (-1)^{2F' + I + J + m_{F}} 
 \sqrt{(2J + 1)(2F + 1)(2F' + 1)} \times
\begin{Bmatrix}
J & J' & 1 \\
F' & F & I
\end{Bmatrix} \notag \\
& \times
\begin{pmatrix}
F' & 1 & F \\
m'_F & m_F - m'_F & -m_F
\end{pmatrix} 
\times
\langle J || e \mathbf{r} || J' \rangle .
\label{eq:Rabi_Frequency}
\end{align}
where, $\langle J || e \mathbf{r} || J' \rangle$ is the reduced dipole matrix element corresponding to the transitions from the ground state $5 \ ^2S_{1/2}$ to the excited state $5 \ ^2P_{3/2}$ of $^{87}$Rb atom. Here $I = 3/2$ is the nuclear spin of $^{87}$Rb atom. The symbol $\Omega_{F,m_{F}}^{F', m'_{F}}$ represents the Rabi frequency for the single-photon absorption for transition $|F, m_F\rangle \rightarrow |F', m'_F\rangle$, while the Rabi frequency $\Omega_{F',m'_{F}}^{F, m_{F}}$ corresponds to the single-photon stimulated emission for the $|F', m'_{F}\rangle \rightarrow |F, m_F\rangle$ transition.

The calculated values of single-photon Rabi frequencies (for total Raman beams power of 9.5 mW) in the $\pi^{0}-\pi^{+}$ two-photon transition routes from $|F = 2, m_F\rangle$ to $|F = 1, m_F\rangle$ via $|F'= 1$ or $2, m'_{F}\rangle$ are shown in the Table \ref{Pi0 - Pi+ transition via F' = 1} and Table \ref{Pi0 - Pi+ transition via F' = 2} for $B_x$ component of the stray magnetic field. Similarly, the Rabi frequencies for $\pi^+$ - $\pi^0$ transitions due to the $B_y$ component and the Rabi frequencies for $\pi^+$ - $\pi^-$ due to $B_z$ component are presented in Table \ref{Pi+ - Pi0 transition via F' = 1}, Table \ref{Pi+ - Pi0 transition via F' = 2}, Table \ref{Pi+ - Pi- transition via F' = 1} and Table \ref{Pi+ - Pi- transition via F' = 2}, respectively.

Assuming a nearly the same population in each  $|F, m_F\rangle$ state, the resultant Rabi frequency corresponding to different Rabi frequencies for  different $|F, m_F\rangle \rightarrow |F', m'_F\rangle$ transitions can be taken as the root mean square (rms) value of these different frequencies as \cite{theory_of_coherent_atomic_excitation,AJDunningCoherent,degeneracy_effect_Rabi_Oscillation},
\begin{equation}
\Omega_{rms} = \sqrt{\frac{\sum_{} \left(\Omega_{F,m_{F}}^{F', m'_{F}}\right)^2}{N}},
\label{eq:RMS_Rabi_Frequency}
\end{equation}
where sum is over all the different single-photon $|F, m_F\rangle \rightarrow |F', m'_{F}\rangle$ transition. Similarly, the resultant Rabi frequency for $|F', m'_{F}\rangle \rightarrow |F, m_F\rangle$ transitions can also be calculated by taking rms value.

\begin{table*}[h!] 
\caption{Rabi frequencies for $\pi^0$ - $\pi^+$ type two-photon Raman transitions via the intermediate state $F' = 1$ for $B_x$ component of the stray magnetic field.}
\scalebox{0.85}{ 
\renewcommand{\arraystretch}{1.5} 
\begin{tabular}{|c|c|c|c|}
  \hline
   \multicolumn{4}{|c|}{\textbf{Two-photon Raman transition routes via $F' = 1$}} \\
   \hline
   \multicolumn{2}{|c|}{\textbf{Stimulated absorption}} & \multicolumn{2}{|c|}{\textbf{Stimulated emission}} \\
   \hline
   \textbf{\boldmath{$|F, m_F\rangle$ $\rightarrow$ $|F', m'_F\rangle$}} & \textbf{\boldmath{$|\Omega_{F,m_F}^{F',m'_F}|/2\pi$ (MHz)}} & \textbf{\boldmath{$|F', m'_F\rangle$ $\rightarrow$ $|F, m_F\rangle$}} & \textbf{\boldmath{$|\Omega_{F',m'_F}^{F,m_F}|/2\pi$ (MHz)}} \\
   \hline
   $|F = 2, m_F = -1\rangle$ $\rightarrow$ $|F' = 1, m'_F = -1\rangle$ & 3.15 & $|F' = 1, m'_F = -1\rangle$ $\rightarrow$ $|F = 1, m_F = 0\rangle$ & 5.25 \\
   \hline
   $|F = 2, m_F = 0\rangle$ $\rightarrow$ $|F' = 1, m'_F = 0\rangle$ & 3.63 &  $|F' = 1, m'_F = 0\rangle$ $\rightarrow$ $|F = 1, m_F = -1\rangle$ & 5.25 \\
   \hline
   $|F = 2, m_F = 0\rangle$ $\rightarrow$ $|F' = 1, m'_F = 0 \rangle$ & 3.63  &  $|F' = 1, m'_F = 0\rangle$ $\rightarrow$ $|F = 1, m_F = 1\rangle$ & 5.25 \\
   \hline
   $|F = 2, m_F = 1\rangle$ $\rightarrow$ $|F' = 1, m'_F = 1\rangle$ & 3.15 &  $|F' = 1, m'_F = 1\rangle$ $\rightarrow$ $|F = 1, m_F = 0\rangle$ & 5.25 \\   
   \hline   
\end{tabular}
}
\label{Pi0 - Pi+ transition via F' = 1}
\end{table*}


\begin{table}[h!] 
\caption{Rabi frequencies for $\pi^0$ - $\pi^+$ type two-photon Raman transitions via the intermediate state $F' = 2$ for $B_x$ component of the stray magnetic field.}
\scalebox{0.85}{ 
\renewcommand{\arraystretch}{1.5} 
\begin{tabular}{|c|c|c|c|}
  \hline
   \multicolumn{4}{|c|}{\textbf{Two-photon Raman transition routes via $F' =2$}} \\
   \hline
   \multicolumn{2}{|c|}{\textbf{Stimulated absorption}} & \multicolumn{2}{|c|}{\textbf{Stimulated emission}} \\
   \hline
   \textbf{\boldmath{$|F, m_F\rangle$ $\rightarrow$ $|F', m'_F\rangle$}} & \textbf{\boldmath{$|\Omega_{F,m_F}^{F',m'_F}|/2\pi$ (MHz)}} & \textbf{\boldmath{$|F', m'_F\rangle$ $\rightarrow$ $|F, m_F\rangle$}} & \textbf{\boldmath{$|\Omega_{F',m'_F}^{F,m_F}|/2\pi$ (MHz)}} \\
   \hline
   $|F = 2, m_F = -2\rangle$ $\rightarrow$ $|F' = 2, m'_F = -2\rangle$ & 8.12 & $|F' = 2, m'_F = -2\rangle$ $\rightarrow$ $|F = 1, m_F = -1\rangle$ & 5.75 \\
   \hline
   $|F = 2, m_F = -1\rangle$ $\rightarrow$ $|F' = 2, m'_F = -1\rangle$ & 4.06 & $|F' = 2, m'_F = -1\rangle$ $\rightarrow$ $|F = 1, m_F = 0\rangle$ & 4.06 \\
   \hline
   $|F = 2, m_F = 1\rangle$ $\rightarrow$ $|F' = 2, m'_F = 1\rangle$ & 4.06 & $|F' = 2, m'_F = 1\rangle$ $\rightarrow$ $|F = 1, m_F = 0\rangle$ & 4.06 \\
   \hline
   $|F = 2, m_F = 2\rangle$ $\rightarrow$ $|F' = 2, m'_F = 2\rangle$ & 8.12 & $|F' = 2, m'_F = 2\rangle$ $\rightarrow$ $|F = 1, m_F = 1\rangle$ & 5.75 \\
   \hline
\end{tabular}
}
\label{Pi0 - Pi+ transition via F' = 2}
\end{table}


\begin{table}[h!] 
\caption{Rabi frequencies for $\pi^+$ - $\pi^0$ type two-photon Raman transitions via the intermediate state $F' = 1$ for $B_y$ component of the stray magnetic field.}
\scalebox{0.85}{ 
\renewcommand{\arraystretch}{1.5} 
\begin{tabular}{|c|c|c|c|}
  \hline
   \multicolumn{4}{|c|}{\textbf{Two-photon Raman transition routes via $F' =2$}} \\
   \hline
   \multicolumn{2}{|c|}{\textbf{Stimulated absorption}} & \multicolumn{2}{|c|}{\textbf{Stimulated emission}} \\
   \hline
   \textbf{\boldmath{$|F, m_F\rangle$ $\rightarrow$ $|F', m'_F\rangle$}} & \textbf{\boldmath{$|\Omega_{F,m_F}^{F',m'_F}|/2\pi$ (MHz)}} & \textbf{\boldmath{$|F', m'_F\rangle$ $\rightarrow$ $|F, m_F\rangle$}} & \textbf{\boldmath{$|\Omega_{F',m'_F}^{F,m_F}|/2\pi$ (MHz)}} \\
   \hline
   $|F = 2, m_F = -2\rangle$ $\rightarrow$ $|F' = 1, m'_F = -1\rangle$ & 4.45 & $|F' = 1, m'_F = -1\rangle$ $\rightarrow$ $|F = 1, m_F = -1\rangle$ & 5.25 \\
   \hline
   $|F = 2, m_F = 0\rangle$ $\rightarrow$ $|F' = 1, m'_F = -1\rangle$ & 1.82 & $|F' = 1, m'_F = -1\rangle$ $\rightarrow$ $|F = 1, m_F = -1\rangle$ & 5.25 \\
   \hline
   $|F = 2, m_F = 0\rangle$ $\rightarrow$ $|F' = 1, m'_F = 1\rangle$ & 1.82 & $|F' = 1, m'_F = 1\rangle$ $\rightarrow$ $|F = 1, m_F = 1\rangle$ & 5.25 \\
   \hline
   $|F = 2, m_F = 2\rangle$ $\rightarrow$ $|F' = 1, m'_F = 1\rangle$ & 4.45 & $|F' = 1, m'_F = 1\rangle$ $\rightarrow$ $|F = 1, m_F = 1\rangle$ & 5.25 \\
   \hline
\end{tabular}
}
\label{Pi+ - Pi0 transition via F' = 1}
\end{table}


\begin{table}[t!]
\caption{Rabi frequencies for $\pi^+$ - $\pi^0$ type two-photon Raman transitions via the intermediate state $F' = 2$ for $B_y$ component of the stray magnetic field.}
\scalebox{0.85}{ 
\renewcommand{\arraystretch}{1.5} 
\begin{tabular}{|c|c|c|c|}
  \hline
   \multicolumn{4}{|c|}{\textbf{Two-photon Raman transition routes via $F' =2$}} \\
   \hline
   \multicolumn{2}{|c|}{\textbf{Stimulated absorption}} & \multicolumn{2}{|c|}{\textbf{Stimulated emission}} \\
   \hline
   \textbf{\boldmath{$|F, m_F\rangle$ $\rightarrow$ $|F', m'_F\rangle$}} & \textbf{\boldmath{$|\Omega_{F,m_F}^{F',m'_F}|/2\pi$ (MHz)}} & \textbf{\boldmath{$|F', m'_F\rangle$ $\rightarrow$ $|F, m_F\rangle$}} & \textbf{\boldmath{$|\Omega_{F',m'_F}^{F,m_F}|/2\pi$ (MHz)}} \\
   \hline
   $|F = 2, m_F = -2\rangle$ $\rightarrow$ $|F' = 2, m'_F = -1\rangle$ & 5.75 & $|F' = 2, m'_F = -1\rangle$ $\rightarrow$ $|F = 1, m_F = -1\rangle$ & 4.06 \\
   \hline
   $|F = 2, m_F = -1\rangle$ $\rightarrow$ $|F' = 2, m'_F = 0\rangle$ & 7.04 & $|F' = 2, m'_F = 0\rangle$ $\rightarrow$ $|F = 1, m_F = 0\rangle$ & 4.69 \\
   \hline
   $|F = 2, m_F = 0\rangle$ $\rightarrow$ $|F' = 2, m'_F = -1\rangle$ & 7.04 & $|F' = 2, m'_F = -1\rangle$ $\rightarrow$ $|F = 1, m_F = -1\rangle$ & 4.06 \\
   \hline
   $|F = 2, m_F = 0\rangle$ $\rightarrow$ $|F' = 2, m'_F = 1\rangle$ & 7.04 & $|F' = 2, m'_F = 1\rangle$ $\rightarrow$ $|F = 1, m_F = 1\rangle$ & 4.06 \\
   \hline
   $|F = 2, m_F = 1\rangle$ $\rightarrow$ $|F' = 2, m'_F = 0\rangle$ & 7.04 & $|F' = 2, m'_F = 0\rangle$ $\rightarrow$ $|F = 1, m_F = 0\rangle$ & 4.69 \\
   \hline
   $|F = 2, m_F = 2\rangle$ $\rightarrow$ $|F' = 2, m'_F = 1\rangle$ & 5.75 & $|F' = 2, m'_F = 1\rangle$ $\rightarrow$ $|F = 1, m_F = 1\rangle$ & 4.06 \\
   \hline
\end{tabular}
}
\label{Pi+ - Pi0 transition via F' = 2}
\end{table}

\begin{table}[h!] 
\caption{Rabi frequencies for $\pi^+$ - $\pi^-$ type two-photon Raman transitions via the intermediate state $F' = 1$ for $B_z$ component of the stray magnetic field.}
\scalebox{0.85}{
\renewcommand{\arraystretch}{1.5} 
\begin{tabular}{|c|c|c|c|}
  \hline
   \multicolumn{4}{|c|}{\textbf{Two-photon Raman transition routes via $F' = 1$}} \\
   \hline
   \multicolumn{2}{|c|}{\textbf{Stimulated absorption}} & \multicolumn{2}{|c|}{\textbf{Stimulated emission}} \\
   \hline
   \textbf{\boldmath{$|F, m_F\rangle$ $\rightarrow$ $|F', m'_F\rangle$}} & \textbf{\boldmath{$|\Omega_{F,m_F}^{F',m'_F}|/2\pi$ (MHz)}} & \textbf{\boldmath{$|F', m'_F\rangle$ $\rightarrow$ $|F, m_F\rangle$}} & \textbf{\boldmath{$|\Omega_{F',m'_F}^{F,m_F}|/2\pi$ (MHz)}} \\
   \hline
   $|F = 2, m_F = -1\rangle$ $\rightarrow$ $|F' = 1, m'_F = 0\rangle$ & 3.15 & $|F' = 1, m'_F = 0\rangle$ $\rightarrow$ $|F = 1, m_F = -1\rangle$ & 5.25 \\
   \hline
   $|F = 2, m_F = 0\rangle$ $\rightarrow$ $|F' = 1, m'_F = -1\rangle$ & 1.82 &  $|F' = 1, m'_F = -1\rangle$ $\rightarrow$ $|F = 1, m_F = 0\rangle$ & 5.25 \\
   \hline
   $|F = 2, m_F = 0\rangle$ $\rightarrow$ $|F' = 1, m'_F = +1\rangle$ & 1.82  &  $|F' = 1, m'_F = +1\rangle$ $\rightarrow$ $|F = 1, m_F = 0\rangle$ & 5.25 \\
   \hline
   $|F = 2, m_F = 1\rangle$ $\rightarrow$ $|F' = 1, m'_F = 0\rangle$ & 3.15 &  $|F' = 1, m'_F = 0\rangle$ $\rightarrow$ $|F = 1, m_F = 1\rangle$ & 5.25 \\   
   \hline   
\end{tabular}
}
\label{Pi+ - Pi- transition via F' = 1}
\end{table}

\begin{table}[t] 
\caption{Rabi frequencies for $\pi^+$ - $\pi^-$ type two-photon Raman transitions via the intermediate state $F' = 2$ for $B_z$ component of the stray magnetic field.}
\scalebox{0.85}{ 
\renewcommand{\arraystretch}{1.5} 
\begin{tabular}{|c|c|c|c|}
  \hline
   \multicolumn{4}{|c|}{\textbf{Two-photon Raman transition routes via $F' =2$}} \\
   \hline
   \multicolumn{2}{|c|}{\textbf{Stimulated absorption}} & \multicolumn{2}{|c|}{\textbf{Stimulated emission}} \\
   \hline
   \textbf{\boldmath{$|F, m_F\rangle$ $\rightarrow$ $|F', m'_F\rangle$}} & \textbf{\boldmath{$|\Omega_{F,m_F}^{F',m'_F}|/2\pi$ (MHz)}} & \textbf{\boldmath{$|F', m'_F\rangle$ $\rightarrow$ $|F, m_F\rangle$}} & \textbf{\boldmath{$|\Omega_{F',m'_F}^{F,m_F}|/2\pi$ (MHz)}} \\
   \hline
   $|F = 2, m_F = -1\rangle$ $\rightarrow$ $|F' = 2, m'_F = -2\rangle$ & 5.75 & $|F' = 2, m'_F = -2\rangle$ $\rightarrow$ $|F = 1, m_F = -1\rangle$ & 5.75 \\
   \hline
   $|F = 2, m_F = -1\rangle$ $\rightarrow$ $|F' = 2, m'_F = 0\rangle$ & 7.04 & $|F' = 2, m'_F = 0\rangle$ $\rightarrow$ $|F = 1, m_F = -1\rangle$ & 2.34 \\
   \hline
   $|F = 2, m_F = 0\rangle$ $\rightarrow$ $|F' = 2, m'_F = -1\rangle$ & 7.04 & $|F' = 2, m'_F = -1\rangle$ $\rightarrow$ $|F = 1, m_F = 0\rangle$ & 4.06 \\
   \hline
   $|F = 2, m_F = 0\rangle$ $\rightarrow$ $|F' = 2, m'_F = 1\rangle$ & 7.04 & $|F' = 2, m'_F = 1\rangle$ $\rightarrow$ $|F = 1, m_F = 0\rangle$ & 4.06 \\
   \hline
   $|F = 2, m_F = 1\rangle$ $\rightarrow$ $|F' = 2, m'_F = 0\rangle$ & 7.04 & $|F' = 2, m'_F = 0\rangle$ $\rightarrow$ $|F = 1, m_F = 1\rangle$ & 2.34 \\
   \hline
   $|F = 2, m_F = 1\rangle$ $\rightarrow$ $|F' = 2, m'_F = 2\rangle$ & 5.75 & $|F' = 2, m'_F = 2\rangle$ $\rightarrow$ $|F = 1, m_F = 1\rangle$ & 5.75 \\
   \hline
\end{tabular}
}
\label{Pi+ - Pi- transition via F' = 2}
\end{table}


\end{document}